\documentclass{IEEEtran4PSCC}
\usepackage{cite}
\ifCLASSINFOpdf
 \usepackage[pdftex]{graphicx}
\else
\fi
\usepackage[cmex10]{amsmath}
\usepackage{multirow}
\usepackage{amssymb}% http://ctan.org/pkg/amssymb
\usepackage{pifont}% http://ctan.org/pkg/pifont
\newcommand{\cmark}{\ding{51}}%
\newcommand{\xmark}{\ding{55}}%

\begin{document}
\title{Numerical Analysis of National Travel Data to Assess the Impact of UK Fleet Electrification}
%Using Travel Survey Data to Analyse the Future Electric Vehicle Fleet in the UK}

%% To specify the authors when (number of affiliations <= 2)
\author{
\IEEEauthorblockN{Constance Crozier, Dimitra Apostolopoulou, Malcolm McCulloch}
\IEEEauthorblockA{Department of Engineering Science \\
University of Oxford\\
Oxford OX1 3PJ, UK\\
\{constance.crozier, dimitra.apostolopoulo, malcolm.mcculloch\}@eng.ox.ac.uk}}
%\and
%\IEEEauthorblockN{Author n.1 Name per Affiliation B\\ Author n.2 Name per Affiliation B}
%\IEEEauthorblockA{(Affiliation B) Department Name of Organization \\
%Name of the organization, acronyms acceptable\\
%City, Country\\
%\{email author n.1, email author n.2\}@domain (if desired)}

%% To specify the authors when (number of affiliations > 2)
% \author{\IEEEauthorblockN{Author n.1\IEEEauthorrefmark{1},
% Author n.2\IEEEauthorrefmark{2},
% Author n.3\IEEEauthorrefmark{3}, 
% Author n.4\IEEEauthorrefmark{3} and
% Author n.5\IEEEauthorrefmark{4}}
% \IEEEauthorblockA{\IEEEauthorrefmark{1} Department Name of Organization A\\
% Name of the organization A,
% Address A\\ Emails if wanted}
% \IEEEauthorblockA{\IEEEauthorrefmark{2} Department Name of Organization B\\
% Name of the organization B,
% Address B\\ Emails if wanted}
% \IEEEauthorblockA{\IEEEauthorrefmark{3} Department Name of Organization C\\
% Name of the organization C,
% Address C\\ Emails if wanted}
% \IEEEauthorblockA{\IEEEauthorrefmark{4}Department Name of Organization D\\
% Name of the organization D,
% Address D\\ Emails if wanted}
% }

\maketitle
\begin{abstract}
Accurately predicting the future power demand of electric vehicles is important for developing policy and industrial strategy. Here we propose a method to create a representative set of electricity demand profiles using survey data from conventional vehicles. This is achieved by developing a model which maps journey and vehicle parameters to an energy consumption, and applying it individually to the entire data set. As a case study the National Travel Survey was used to create a set of profiles representing an entirely electric UK fleet of vehicles. This allowed prediction of the required electricity demand and sizing of the necessary vehicle batteries. Also, by inferring location information from the data, the effectiveness of various charging strategies was assessed. These results will be useful in both National planning, and as the inputs to further research on the impact of electric vehicles.
\end{abstract}


% Use this to place sponsorships
%\thanksto{Applicable sponsors, if any, should be placed using the \emph{thanksto} command}

\section{Introduction}

\par{In the coming years, a large increase in the number of electric vehicles (EVs) on the UK roads is expected; predictions suggest that by 2040 more than half vehicles purchased will be electric\cite{bloomberg}. This could be key in helping the government achieve its reduced carbon and emissions targets as the electricity used to power EVs can be produced from renewable sources. However the mass adoption of EVs is likely to present some challenges; substantial charging infrastructure will need to be built and vehicle charging will significantly change the power demand profile.} \\

%\par{Currently less than 4\% of registered vehicles in the UK are pure electric, however this is more than double the number there were two years ago, so a large electric fleet could only be a few years off.\cite{ev-stats} In order to successfully plan for such a scenario, the amount of energy required by the vehicles must be accurately predicted.
% I need some more motivation here

%\par{Research into this area is a hot topic, in the UK several EV trials have been conducted, such as \textit{My Electric Avenue}, where the actual energy consumption of the vehicles is recorded. However, the small number of participants results in small and, most likely, skewed datasets. This paper instead proposes using survey data detailing the behaviour of conventional vehicles, assuming that electrification will not change driver behaviour.}

%\par{In order to evaluate the scale of the problem, and the potential effect of various charging strategies, the amount of energy required by the vehicles must be accurately predicted. Several trials have been conducted where electric vehicles were given to households, and their consumption and activity logged. An example of this is the \textit{My Electric Avenue} project, where Nissan Leafs were lent to 150?? households on the same network for a period spanning 3 years. While the data collected from such studies is high quality, it represents a very small, and likely biased, sample. National travel surveys on the other hand

\par{Various work has been carried out investigating the capacity\cite{Darabi2011}, infrastructure\cite{PieltainFernandez2011} and economic\cite{Vagropoulos2013} impacts of an increased electric vehicle fleet. All of these studies require as inputs a model for individual vehicles' energy demand. The results will likely influence government policy and industrial strategy, so it is important that the input data accurately reflects the population in question. The top two concerns of potential EV purchasers has found to be the limited range and the ability to charge\cite{Bonges2016}, so obtaining a better understanding of likely charging requirements may also accelerate the uptake of EVs.}

\par{Although some trials have been carried out investigating the real consumption of electric vehicles (e.g. \cite{mea}), the small scale mean that they are not necessarily representative of the behaviour of a large electric fleet. However, by assuming that electrification does not significantly alter consumer behaviour, conventional vehicle data can be used to analyse the future behaviour of electric vehicles. This approach is advantageous because data concerning conventional vehicles has been gathered for many years, resulting in large and extensive data sets. This assumption has been previously exploited; \cite{Huang2010} and \cite{Bucher2015} use UK travel survey data as the basis for Monte Carlo Simulations, forming discrete probability distributions from the aggregated statistics. In \cite{Huang2010} distributions for number of vehicles in a household, home arrival time and distance travelled are constructed then sequentially sampled, while \cite{Bucher2015} assumes distributions for number of journeys undertaken, purpose of journey and length of journey.}

\par{These models are stochastic, allowing the incorporation of uncertainty and an unlimited number of profiles can be generated. However, by just using the aggregated results, information about the individual users is lost. The simulations will produce realistic results on large scales (i.e. when looking at the aggregated demand of a group of EVs), but there is no evidence to suggest they capture representative individual vehicle behaviour.} \\

%\par{This paper proposes using a large travel survey data set to create a deterministic bottom-up model for charging demand in the UK. By individually modelling vehicles' theoretic electricity consumption, we aim to accurately capture the behaviour of individual vehicles. Predictions for the geographic and temporal distribution of energy required can then be made, and the effectiveness various charging strategies assessed.}

%\par{The rest of this paper is organised as follows: Section \ref{sec:NTS} introduces the data set used as a basis the simulations, section \ref{sec:vehicle-model} introduces the modelling techniques necessary to convert the travel data into energy predictions, section \ref{sec:methodology} outlines the method for running a simulation, and section \ref{sec:res} interprets the results, considering: the change in consumption with time, the required size of vehicle batteries and the effective placing of charging stations.}

\par{This paper proposes a methodology for creating a representative set of energy demand profiles using travel survey data. A vehicle energy model is developed which allows different makes of vehicle and number of passengers to be incorporated into the calculations. As a case study the UK fleet is considered, in the case where 100\% of vehicles are EVs. The resulting data set is used to size the required vehicle batteries and quantify the likely power demand. Also, by inferring location information from the data, the effectiveness of various charging strategies can be assessed.}

\section{Data} \label{sec:NTS}

\par{The National Travel Survey (NTS) is an annual survey, which aims to determine how people in the UK use transport. Households are selected at random and asked to document all of their journeys for the week, recording (among other things) their day, time, distance, length, purpose and mode of transport. Regional and demographic data for the participating households is also collected. Diaries from \(91,755\) households owning at least one vehicle are available, com prising a total of \(1,862,168\) trips.\cite{nts}}

\par{Other countries carry out similar travel research, for example in the US they have the \textit{National Household Travel Survey}. The advantage of using data from such surveys to predict vehicle usage is that their low cost results in large sample sizes, which should represent a broad range of population.}

%While many vehicle trials have been carried out to analyse travel patterns (eg. REFRENCE), these normally have a small and potentially biased group of participants. The disadvantage of the travel survey approach is that the data is recorded by participants, often in retrospect, so the information may not be reliable.}

\section{Vehicle Model} \label{sec:vehicle-model}

\par{In order to predict the energy expenditure by a vehicle on a journey a model is required. Other studies, \cite{Huang2010} and \cite{Bucher2015}, assume a direct proportionality between energy consumption and distance travelled, however this is very simplistic. \cite{Wu2011} creates a stochastic model which randomly assigns a proportionality coefficient conditioned on vehicle type (car, var or SUV), which introduces diversity into the simulation but does not account for variation in region or weight. Heavy vehicles in urban areas will use more energy than a light vehicle on the motorway, and the NTS data includes both the number of passengers and the type of region in which a journey took place.}

\par{While this model should incorporate varying mass, regional characteristics and vehicle make, it must also be computationally inexpensive as it will be applied in the order of \(10^6\) times per simulation. Here we propose a \textit{tank-to-wheel} model, which maps a drive cycle to an energy consumption.}

\subsection{Drive cycles}

\par{A drive cycle is essentially a trace of a vehicle's velocity during a journey. If GPS data is available it is possible to recover a closely representative drive cycle for each recorded journey, however we only have survey data. It is therefore necessary to find a set of standard drive cycles which are representative of how people in the UK drive.}

\par{The ARTEMIS project analysed actual data of European car driving and produced a set of real-world standard drive cycles\cite{artemis}. Three different cycles were produced, representing urban, rural and motorway driving conditions respectively.}

\par{For each journey in the dataset a drive cycle is formulated by looping the relevant ARTEMIS cycle until the recorded distance is achieved. For journeys over 10 miles the motorway cycle is used, otherwise the cycle is chosen based on the rural/urban classification of the household.}

\par{If this analysis were repeated for a non-European country a different set of, more representative, drive cycles should be used.}

\subsection{Tank-to-wheel model}

\par{In order to calculate the total energy used it's useful to consider the power required by the wheel-axle at each time-step, \(P_t\). This is given by the product: 
\begin{align} \label{eq:pfv}
P_t = F_t v_t ,
\end{align}
where \(v_t\) is the vehicle's speed and \(F_t\) is the force which it must overcome at time \(t\). This force can be decomposed into:
\begin{align} \label{eq:force}
F_t = F_t^{res} + m a_t ,
\end{align}
where \(m\) is the total mass of the car and load, \(a_t\) is the acceleration at that time-step and \(F_t^{res}\) is the resistive force on the vehicle. Resistive force is commonly broken down into a constant term, one proportional to velocity and one to velocity squared\cite{elements-of-mechanics}. It should be noted that in reality these terms vary with road and weather conditions, but as we don't have information about these quantities it would be difficult to incorporate them into the model. The total force is therefore described as:
\begin{align} \label{eq:resistance}
F_t^{res} = f_0 + f_1 v_t + f_2 v_t^2 ,
\end{align}
where \(f_0\), \(f_1\) and \(f_2\) are vehicle specific coefficients, dependant on characteristics such as frontal area and drag coefficient. Empirical values for these can be determined from a coast-down test, where a vehicle is allowed to decelerate with the engine off\cite{coast-down}. This is one of the tests which manufacturers are required to carry out and the results for every vehicle available in the US are published by the Environmental Protection Agency (EPA)\cite{epa-data}.
By substituting \eqref{eq:resistance} and \eqref{eq:force} into \eqref{eq:pfv} and utilising a backwards-difference approximation (e.g.\cite{finite-difference}) for the acceleration term we arrive at:
\begin{align}
P_t =  f_0 v_t + f_1 v_t^2 + f_2 v_t^3 + m \frac{v_t-v_{t-1}}{\Delta t} v_t ,
\end{align}
where \(\Delta t \) is the size of the time-step. This is the power required by the wheel-axle, not the engine; there will be some losses in the engine which must be accounted for. The amount of power lost will depend largely on the vehicle, but also on the drive cycle and external conditions. Here we make the simplifying assumption that the power required from the tank at time \(t\), \(P_t^{req}\), is given by:
\begin{align}
P_t^{req} = 
\begin{cases}
  P_0 + \eta \text{ } P_t \text{ } \text{ for } a_t < 0 \\    
  P_0 + \frac{1}{\eta} \text{ } P_t \text{ } \text{ for } a_t > 0  
\end{cases}
\end{align}
where \(P_0\) is a constant power loss \(\eta\) is an efficiency specific to the vehicle model. This is a large simplification however by learning this value from real-world data un-modelled complexities can be \textit{absorbed}. For example, electric vehicles will often use regenerative braking, which isn't modelled here, but this will be reflected by higher values of \(\eta\). The total energy expenditure requirement, \(E\), of a journey will then be given by the area under the power requirement curve, such that:
\begin{align}
E = \sum_{t =1}^{T} {P_t^{req}} \Delta t .
\end{align}}

\subsection{Learning the Parameters}

\par{As well as publishing the mass and coast-down coefficients of all EVs currently for sale the EPA also records the energy consumption of the vehicles when put through a set of standard drive cycles on a dynamometer. For electric vehicles two test cycles are recorded: one representative of urban driving and one for highways driving. If the dataset includes \(n\) vehicles this results in \(2n\) data points. We have \(n+1\) parameters to learn - one efficiency per vehicle and a constant loss term, so providing data for more than one vehicle is available a set of best-fit parameters can be learnt by minimising the mean square-error of the estimated energy consumption.}

\par
\begin{figure} 
\centering
\includegraphics[width=8cm]{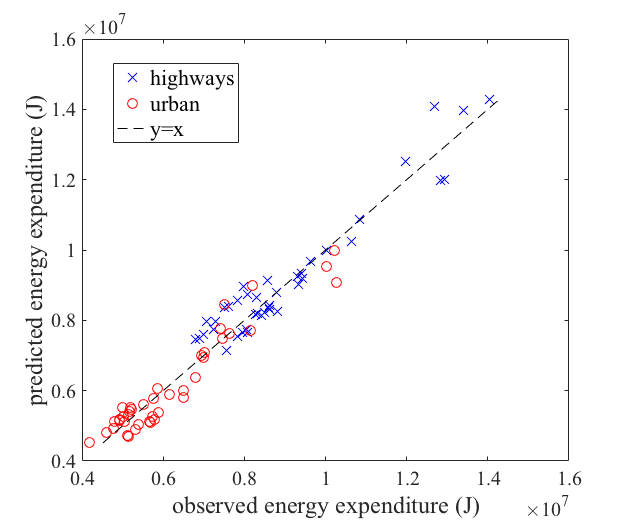}
\caption{A quantile-quantile plot of the vehicle model performance on the two recorded drive cycles}
\label{fig:model-performance}
\end{figure}

{The performance of the resulting model is shown using a quantile-quantile plot in Figure \ref{fig:model-performance} . There were 41 models of electric vehicle and each contributes two points to the plot - one for each drive cycle. The average error in the predictions was 5.2\% for the highway and 5.7\% for the urban drivecycle. This error could be reduced by adding more expressivity to the model, for example by letting \(P_0\) vary between vehicles. However, given the small amount of data available there is a trade-off between model performance and reliability.} \\

\par{For the rest of the paper the Nissan Leaf model is used exclusively. While there are other EVs available in the UK this is the most common, and no vehicle information is recorded in the travel survey so there would be no basis to distinguish between vehicle types. However, later a more diverse fleet could easily be incorporated. The parameters which define this model are listed in Table \ref{tab:leaf-model}, where the first 4 parameters are taken from the EPA data directly and the second two are learnt from the sample energy expenditures.}

\begin{table}[]
\centering
\caption{Nissan Leaf Model Parameters}
\label{tab:leaf-model}
\begin{tabular}{|c|c|c|c|c|c|}
\hline
m (kg) & \(f_0\) (N) & \(f_1\) (N\(sm^{-1}\)) & \(f_2\) (N\(s^2m^{-2}\)) & \(P_0\) (kW)     & \(\eta\)  \\ \hline
1521    & 133 & 0.756                         & 0.489                                             & 1.17 & 86.0\% \\ \hline
\end{tabular}
\end{table}

\par{As well as the energy used to complete the drive cycle an electric vehicle also has an \textit{accessory load}, which is the power consumption due to onboard electronics such as lights, radio and the heater. Real vehicle energy consumption was examined and the accessory load was found to be most strongly dependant on time of year, likely due to the dominance of the heating in the power consumption. Therefore a month-specific constant accessory load was assumed, with values varying from 0.2 to 1.6 kW.}

\section{Methodology} \label{sec:methodology}

\par{In order to run a simulation of the energy demanded by a fleet in a certain period of time the following algorithm was used to process the whole data set:
\begin{enumerate}
  \item Filter the data set 
  \item Assign trips to vehicles
  \item Convert trips into a predicted energy expenditure
  \item Aggregate each vehicles predicted consumption
\end{enumerate}}

\par{In the first step the trip file is filtered to select only journeys suited to the current simulation. Common traits to filter for include day of the week, month, region and region type. UK travel patterns have not changed significantly in the last 15 years,\cite{nts} so the survey year of trips was ignored in order to increase the size of the useable data.}

\par
\begin{figure}
\centering
\includegraphics[width=8cm]{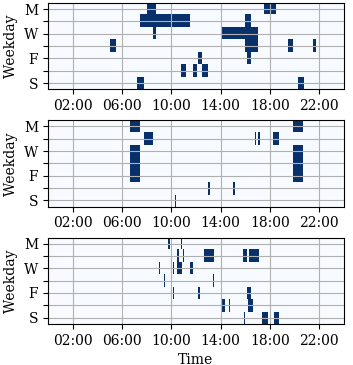}
\caption{Three example one-week vehicle usage profiles. Shaded areas indicate times the vehicles are in use}
\label{fig:usage-profiles}
\end{figure}

{Next trips must be assembled into travel diaries from the vehicles' perspective, which can be up to a week long (the length of the survey period). Example usage profiles for three vehicles from the data set are shown in Figure \ref{fig:usage-profiles}, where day of the week is plotted on the vertical axis, time of day on horizontal and shaded areas indicate that the vehicle is in use. You can see that the second vehicle has almost identical journeys on 4 of the days, possibly a commute. Also note that the first vehicle is prone to much longer journeys than the third. This type of agent behaviour would not be captured using a top-down stochastic approach.}

%\par{As well as recovering the individual profiles, the total number of vehicles represented by the remaining data should be calculated.}

\par{The journeys are defined by a distance, start time, length, number of passengers and region type. Each journey is run through the vehicle model described in Section III, assuming each of the passengers weighs 70kg.\cite{bodyweight} These results are then summed for each vehicle to estimate the energy consumption of that vehicle in the simulation period.}

\section{Results and Discussion} \label{sec:res}

\par{Several simulations were carried out, investigating the behaviour of the hypothetical UK electric fleet under various parameters.}

% I might lose the stuff below:
\par
\begin{figure} 
\centering
\includegraphics[width=8.5cm]{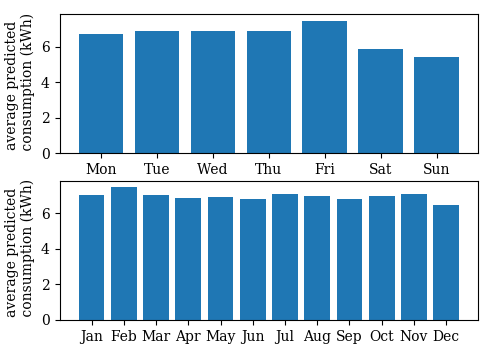}
\caption{The temporal variation in predicted daily energy consumption}
\label{fig:seasonal-variation}
\end{figure}

{Figure \ref{fig:seasonal-variation} shows the variation in average daily vehicle predicted energy consumption with both weekday and month. The variation in energy prediction throughout the week and year is significantly less than the variation in number of journeys or mileage, as on days/months where fewer journeys are undertaken (such as August or Sunday) trips are likely to be longer and vehicles fuller. It should be noted that the weekday variation is more significant than the month; travel diaries were a week long so while the weekday variation represents the behaviour of the same vehicles, the month variation does not.}

\par
\begin{figure} 
\centering
\includegraphics[width=8cm]{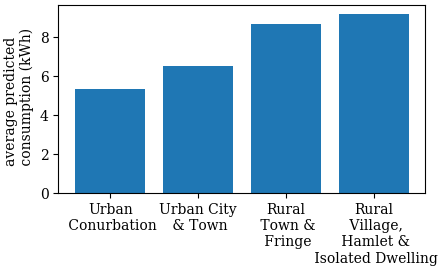}
\caption{The variation in consumption with rural-urban classification}
\label{fig:rural-variation}
\end{figure}

{Figure \ref{fig:rural-variation} shows the variation in predicted consumption with rural-urban classification. The four categories used are defined by the Office for National Statistics\cite{rural-urban} and are positioned in order of increasing sparsity. This shows that, on average, the more rural a household the higher their vehicular energy consumption.}

\subsection{Vehicle Location}

\par{Charging a vehicle from empty can take over 12 hours so in order to design charging strategies it is important to know where vehicles are parked. It is possible to infer this information from the travel survey trip purposes, for example if a car travels to work we can assume it will be parked at work until its next journey. Unfortunately there is no way to further breakdown this information - for example to determine which of the vehicles are likely to be parked in office complexes.}

\par
\begin{figure} 
\centering
\includegraphics[width=8.5cm]{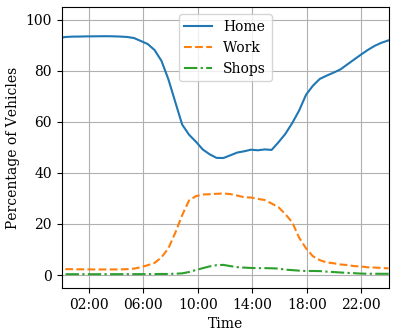}
\caption{The fleet location with time on a weekday}
\label{fig:vehicle-location}
\end{figure}

{Figure \ref{fig:vehicle-location} shows the predicted location of the fleet grouped into location types; work and shopping centres were chosen as these are commonly suggested charging locations. According to this simulation more than 50\% of vehicles are at home at any one time, suggesting that home would be the most convenient place to charge. There are also a significant number of vehicles at work during the middle of the day, while only a modest number of vehicles appear to be parked at the shops.}

\subsection{Energy Demand}
\par
\begin{figure} 
\centering
\includegraphics[width=9cm]{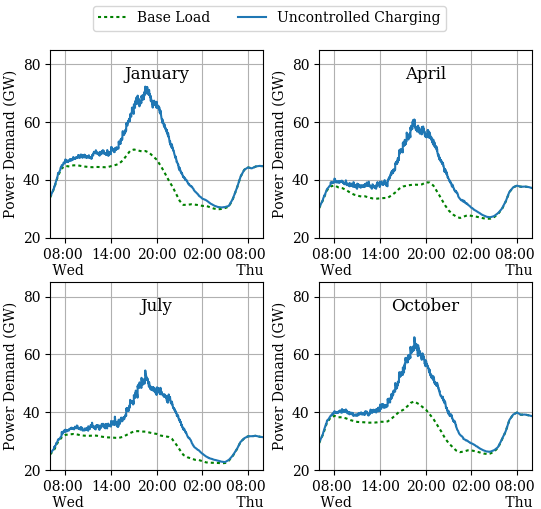}
\caption{The predicted impact of charging an electric fleet on the demand profile in the UK throughout the year}
\label{fig:national-profiles}
\end{figure}

{By assuming that all users plugged their vehicle in as soon as they got home from their last journey and charged it until full, a power demand profile can be constructed. Figure \ref{fig:national-profiles} shows the UK base power demand compared to the predicted total demand with a 100\% electric fleet. This simulation was carried out on a Wednesday in January, April, July and October respectively.}

\par{The predictions show a significant increase in the peak demand; largely because the peaks of the domestic and charging load are both when people arrive home. This suggests that if uncontrolled at home charging were to be accommodated an additional 20GW of generation capacity would be required. The increase in variation throughout the day is also likely to present some problems, as it makes accurate demand prediction more difficult. The difference between months is dominated by the base load, which varies throughout the year due to heating. It should be noted that although the peak in July is the lowest, power stations are often switched off for refurbishment during the summer, reducing the available capacity.}

\subsection{Battery Sizing}
\par{Another use of the model developed is in battery sizing; by predicting the amount of energy a vehicle will use a suitable battery capacity can be chosen.}

\par{To begin with it was assumed that each car would charge exactly once per day, as this is the strategy advertised by manufacturers. This means a vehicle would need the capacity to complete all journeys within that day, but no extra. Table \ref{tab:dailyconsumption} shows the predicted daily consumption of all of the vehicles in the data set binned into discrete intervals.}

\begin{table}[]
\centering
\caption{The distribution of predicted daily energy consumption in kwh}
\label{tab:dailyconsumption}
\begin{tabular}{|c|c|c|c|c|c|c|c|}
\hline
0      & 0-5  & 5-10 & 10-20 & 20-30 & 30-50 & 50-100& \textgreater 100 \\ \hline
27.0\% & 38.6\% & 17.6\% & 9.0\%   & 3.8\%   & 2.4\%   & 1.4\%    & 0.2\%            \\ \hline
\end{tabular}
\end{table}
\par{More than a quarter of vehicles were not used at all on the chosen day, and the majority of those which were used less than 5 kWh. This suggests that the public perception that electric vehicles have insufficient range is unwarranted - in fact, 94.1\% of vehicles used less than 24 kWh which is the capacity of a Nissan Leaf. EVs with larger batteries are currently available, however these are much heavier and in the energy calculations the mass of a Nissan Leaf was assumed, so for a vehicle with a larger battery more energy than predicted would be required}

\par{Although these figures are useful for analysing vehicle usage it is naive to size a vehicle's required battery from just one day of use. A better method would be to look at the maximum energy expenditure in any day during the recorded period, in other words the required size of battery to complete the travel diary with a maximum of one charge a day. This information is listed in Table \ref{tab:batterysize}.}
\begin{table}[]
\centering
\caption{The distribution of predicted required battery size in kWh to complete the recorded week of journeys with 1 charge per day}
\label{tab:batterysize}
\begin{tabular}{|c|c|c|c|c|c|c|}
\hline
0-5  & 5-10 & 10-20 & 20-30 & 30-50 & 50-100 & \textgreater 100 \\ \hline
32.7\% & 21.2\% & 18.7\% & 10.5\%   & 9.2\%   & 6.6\%   & 1.2\%             \\ \hline
\end{tabular}
\end{table}

\par{As expected this distribution carries more weight at the larger capacities, and no vehicles were unused - if a vehicle wasn't used all week it wouldn't be part of the dataset. This simulation suggested that for 77.5\% of vehicles in the dataset a Nissan Leaf would have sufficient capacity with a maximum of one charge per day.}
 
\subsection{Charging Strategies}

\par{Of those vehicles in the single day simulation which exceeded the Nissan Leaf capacity it is possible that many could have avoided running out of charge by charging in between journeys. The possibility of mid-day charging depends on both charging station location and vehicle availability; if a vehicle is in use all day, or never parked near a station then recharging would not be possible.}

\par{By considering the locations of the vehicles which would have run out of charge we can consider the effectiveness of these charging locations in allowing mid-day charging. The most basic solution is to allow mid-day charging at home, as the ability to charge at home has already been assumed. More flexible scenarios include the opportunity to charge at work or at the shops, though these would require additional infrastructure. For a single day simulation the percentage of vehicles exceeding their capacity was estimated using a combination of charging strategies, and the results are displayed in Table \ref{tab:charging-schemes}.}

\begin{table}[]
\centering
\caption{The effectiveness of different locations for mid-day charging}
\label{tab:charging-schemes}
\begin{tabular}{|c|ccc|c|}
\hline
\multirow{2}{*}{Scenario} & \multicolumn{3}{c|}{Mid day Charging}                         & \multirow{2}{*}{\% Out of charge} \\
                          & \multicolumn{1}{c|}{home} & \multicolumn{1}{c|}{work} & shops &                                   \\ \hline
(i)                         & \xmark                         & \xmark                         & \xmark     & 5.90\%                                 \\
(ii)                        & \cmark                         & \xmark                         & \xmark     & 5.23\%                                 \\
(iii)                       & \cmark                         &\cmark                         & \xmark     & 3.84\%                                 \\
(iv)                        & \cmark                         & \xmark                         & \cmark     & 5.08\%                                 \\
(v)                         &  \cmark                        &  \cmark                         &  \cmark     & 3.69\%                                 \\ \hline
\end{tabular}
\end{table}

\par{These results suggest that a modest reduction in over capacity vehicles can be achieved through simply allowing busy vehicles at home to charge in the middle of the day. The most significant further improvement was obtained by incorporating work charging into the scheme, implying a significant proportion of the vehicles covering large distances are commuting. By contrast, allowing vehicles to charge at the shops made little difference, suggesting that either vehicles were not parked at the shops for long enough to receive significant charge, or that vehicles parked in shopping centres are not the ones requiring more than one charge a day.}

\subsection{Comparison with Existing Results}

\par
\begin{figure} 
\centering
\includegraphics[width=8.5cm]{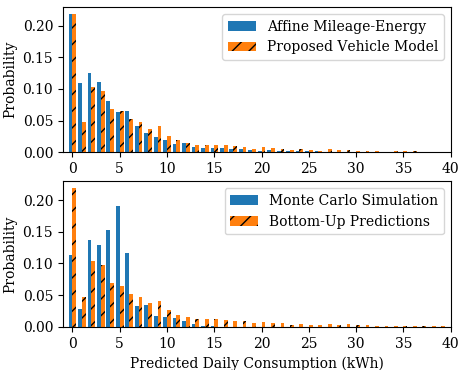}
\caption{A comparison of this papers results with: a) a constant ratio between mileage and consumption, b) a stochastic journey model}
\label{fig:model-comparison}
\end{figure}

{A common approach observed in the literature was to use a affine relationship between the distance travelled and the energy consumed (e.g. \cite{Huang2010}), whereas this paper proposed a higher fidelity vehicle model. The difference in the predicted individual vehicle daily energy consumption with the proposed model, opposed to a constant 0.23 kWh / mile is shown in Figure \ref{fig:model-comparison}(a). This suggests the affine approach results in more conservative estimates, particularly for the vehicles which experienced lighter use; possibly because shorter journeys are likely to be less energy efficient.}\\

\par{In other studies prediction focused on the journeys being completed by a fleet rather than individual vehicles (e.g. \cite{Bucher2015}). For a large enough fleet of vehicles the aggregated energy consumption of the vehicles should be representative, however the smaller the fleet the less confident these predictions can be.}

\par{To demonstrate this a Monte Carlo simulation based on the NTS data was carried out to generate the journeys completed by the UK fleet. Journeys were generated in pairs, comprising an out and a return, and allocated at random to of one the vehicles available for the whole trip duration. The distribution of total energy consumed per vehicle is compared with the bottom-up approach in Figure \ref{fig:model-comparison}(b). The simulation has failed to capture the individual vehicle behaviour, with probability mass being pulled towards an average value. For battery sizing this method would be completely unsuitable; it would appear to suggest that a negligible number of vehicles would use more than half their batteries a day.}

\section{Conclusion}

\par{In this paper we have proposed a low complexity bottom-up model for predicting the vehicular electricity consumption in the UK. The bottom up approach allowed individual vehicles' needs to be captured. This makes the predictions superior to those present in literature; it is clear that journeys are not uniformly distributed between vehicles, with user-specific patterns being evident in the usage profiles.}\\

\par{It was predicted that the majority of vehicles would use less than 5 kWh a day, well less than the battery capacity of a small EV. This suggests that, so-called \textit{range anxiety} is unwarranted.}

\par{Relatively little variation was found in energy consumption with time, although requirements were notably lower at weekends. The rural-urban classification having a much more significant effect on predictions, with the more rural areas requiring more energy per vehicle.}\\

\par{The most convenient location to charge is at home, as this is where vehicles spend most of their time. After this, at work charging appeared to show more promise than shopping centre charging. However, it may be that if charging stations were available at shopping centres consumers would change their behaviour to use them. The potential infeasibility of at home charging, e.g. due to limited off-road parking, has also not been considered.}

\bibliography{references.bib}{}
\bibliographystyle{IEEEtran}

\end{document}